# Ray dynamics of whispering-gallery modes in curved space

YONGSHENG WANG, XIAOXUAN LUO, ZAOYU CHEN, ZHENZHI LIU, YAPING HOU , FU LIU, YIN CAI, YANPENG ZHANG, AND FENG LI[*]

*Key Laboratory for Physical Electronics and Devices of the Ministry of Education & Shaanxi Key Lab of Information Photonic Technique, School of Electronic Science and Engineering, Faculty of Electronic and Information Engineering, Xi'an Jiaotong University, Xi'an 710049, China*
*\* Corresponding author: felix831204@xjtu.edu.cn*

**Abstract:** Microcavities supporting Whispering-gallery modes (WGMs) are of great significance for on-chip optical information processing, which is frequently analyzed using ray dynamics that conventionally involves straight light trajectories in flat space. We develop a novel mathematical model of ray dynamics that allows investigating photon movement in WGM-microcavities defined on curved surfaces, which consists of curved or twisted light trajectories following the geodesic lines of the space. We show that the resulting ray dynamics differs dramatically from those in flat microcavities, and in various manners depending on whether the curved surface can be unfolded to flat. Our methods suggest a redefinition of the WGM symmetry in curved space and provides a universal tool for analyzing three-dimensional (3D) photonic circuits containing curved structures with boundaries.

## 1. Introduction

Optical microcavities supporting whispering gallery mode (WGM) are resonators confining light via successive internal reflections that are extensively investigated for applications in on-chip microlasers [1,2], sensors [3,4], filters [5], isolators [6], frequency combs [7] and gain-loss featured optoelectronic devices [8,9]. The cavity periphery basically exhibits a circular or polygonal shape in forms of microdisks, microspheres, microtoroids, microwires and microtubes [10–15]. To achieve particular features such as unidirectional lasing, asymmetric cavities were designed and fabricated by either abruptly or smoothly deforming the cavity periphery [16,17], which leads to stable and chaotic optical modes [18–20]18-20. It is shown that the chaotic modes in deformed cavities are particularly useful to achieve broadband resonance and nonlinear effects such as frequency combs [21–23]. Among the methods for analyzing optical modes in deformed cavities, ray dynamics is one of the most efficient: light in the cavity are modeled as straight lines reflecting successively at the cavity periphery, and the reflecting angle was recorded as a function of the reflecting position in a diagram called the Poincare surface of section (PSOS) [24–27], which enables the clear identification of WGM, stable and chaotic modes.

Recently, curved membranes are developed to form three-dimensional (3D) on-chip photonic structures, which would allow much higher capacity of devices on integrated photonic circuits compared to the two-dimensional (2D) counterparts [28–32]. Interestingly, it has been experimentally demonstrated that a 2D microdisk cavity can be bent up to support WGMs exhibiting 3D trajectories in space, while keeping a considerably high quality factor[33,34]. The question thereby arises whether curved 2D structures can be described efficiently using ray dynamics. Especially, one would naturally wonder whether a WGM cavity build on a curved 2D space, regardless of the details of its shape, can still be efficiently analyzed using ray dynamics and whether dramatic differences can be found between WGMs in curved and flat spaces. Nevertheless, the traditional ray dynamics assuming straight light trajectories can no longer fit the curved structures, and therefore novel analyzing methods are to be established.

In this article, we propose an efficient model of ray dynamics for the analysis of WGM in curved space. Instead of being modeled as straight, light travels along the geodesic line defined by the shape of the curved 2D space, and reflects at the microcavity boundaries which are generally space curves. We show that in the curved WGM cavities, deformation does not necessarily lead to asymmetry. In addition, the asymmetry is not only presented by the positon and momentum of a photon as in the PSOS, but also by a third degree of freedom: the geometry of the trajectory of the photon, including its curvature and torsion. We also found remarkable differences of light trajectories between WGM cavities in curved and flat spaces. Our results suggest a redefinition of symmetry when an extra degree of freedom is added to WGM microcavities and provide a general method for the efficient analysis of curved photonic structures with boundary reflections using ray dynamics.

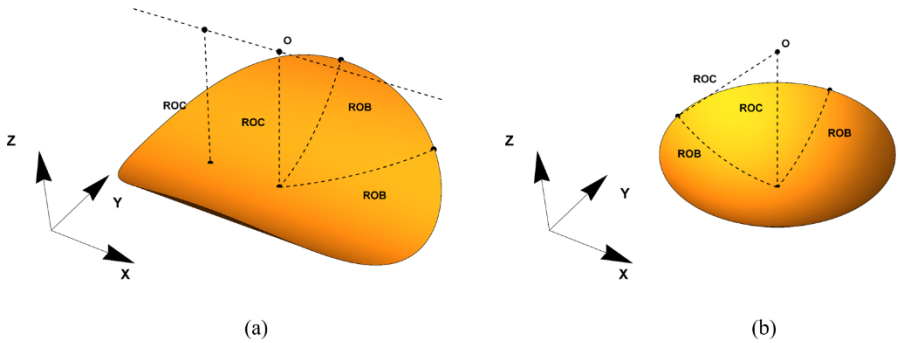

Fig.1. Illustrative pictures of the cylindrically (a) and spherically (b) curved cavities.

## 2. The Model

We start from the relation between asymmetry and deformation. In comparison with flat 2D surfaces, curved 2D surfaces can be classified into two categories: the ones that can (named as Type I hereinafter) and cannot (named as Type II hereinafter) be unfolded to become a flat surface. Typical examples of Type I and Type II WGM cavities are cylindrical and spherical surfaces with circular boundaries respectively, as illustrated in Fig. 1 (a) and (b). Herein "circular" means that each point on the boundary shows equal distance to the geometrical center measured along the geodesic lines of curved surface. Whilst the two cavities are both undoubtedly identified as deformed compared to flat WGM cavities, they are different in terms of symmetry. The cavity of Fig. 1(b) (referred to as spherical cavity hereinafter) is completely symmetric as a curved 2D structure, since it shows the same radius of curvature (ROC) and radius of boundary (ROB) everywhere. Nevertheless, the cavity of Fig.1 (a) (referred to as cylindrical cavity hereinafter) is curved in the Y direction but flat in the X direction, and is thereby not completely symmetric. Indeed, it is symmetric in terms of boundary but asymmetric (herein uniform but anisotropic) in terms of the metric of curvature. We will show that such a difference in symmetry leads to very different behaviors of optical modes inside the cavities.

We regard the curved 2D surfaces as very thin waveguides in which light travels with a uniform effective refractive index $n_{eff}$. Therefore, like travelling in straight lines in flat 2D cavities, light ray travels in geodesic lines in the curved space, which ensures the optical length of extremum it passes. The light trajectory of geodesic line intersects with the cavity boundary, and reflects at the intersecting point back into the cavity following the law of reflection, i.e., equal incident and reflecting angle with the tangent of the boundary. The algorithm thus contains 6 steps:

(1) Choose an arbitrary starting point $P_0$ at the cavity boundary (which is generally a spatial curve C) and an arbitrary initial traveling direction $l_0$ ( a tangent vector of the curved cavity surface Σ at $P_0$) ;
(2) Derive the equation of the geodesic line $S$ with $P_0$ and $l_0$. $S$ is then the trajectory of the incident beam. Calculate the curvature $B$ and torsion $T$ of $S$.
(3) Derive the intersecting point $P$ between the incident beam $S$ and the cavity boundary $C$, by simultaneously solving the equations of both. $P$ is then the reflecting point.
(4) Derive the tangent vector of $S$ at $P$, noted as vector $l$.
(5) Derive the tangent vector of $C$ at $P$. noted as vector $t$. The angle formed by $l$ and $t$ is therefore the complementary angle of the incident angle at $P$.
(6) Derive the tangent vector of the reflected beam at $P$, noted as $l'$, following the law of reflection that the incident and reflecting angles are equivalent (so are their complementary angles), i.e., $l' \cdot t = l \cdot t$. Meanwhile, the reflected light has to be inside the cavity surface Σ, i.e., $l'$ has to be tangent to Σ, which requires $l' \cdot n = 0$, where $n$ is the normal vector of Σ at $P$. For cylindrical and spherical surfaces, $n$ is the unit vector along the radial directions of the sphere and the cylinder, respectively.
(7) Using $P$ and $l'$ as the new starting point and initial direction vector, run steps (2) –(6) repeatedly. After enough times of iterations, choose again a new set of $P_0$ and $l_0$ for a new round of calculation.

In the algorithm, all tangent and normal vectors, as well as the curvatures and torsions can be calculated using standard mathematical tools of vector analysis. Although the derivation of geodesic line equations in general requires tensor analysis (thereby differentially geometry), for the specific situation of regularly shaped surfaces like the cylindrical and spherical ones, existing solutions of mathematical expressions are available, i.e., the helix and the great circle, respectively. In each iteration of the algorithm, the position $P$, direction vector $l$, curvature $B$ and torsion $T$ are recorded. The collection of $l$ vs. $P$ constitutes the PSOS which is the standard analyzing method also for flat 2D cavities, whilst $B$, $T$ vs. $P$ constitutes the diagram of trajectory geometry (DTG) which is a novel concept in our approach. We show that for WGM cavities defined on a curved 2D surface, both the PSOS and the DTG are required in ray dynamics for a comprehensive characterization of the cavity properties.

## 3. Results and Discussion

We model the cylindrical and spherical cavities as typical examples of Type I and Type II cavities, respectively. In each model, we first perform calculations of cavities with symmetric, or "circular" boundaries, and then those with asymmetric, or deformed "circular" boundaries. Finally, we compare the calculated results with the flat counterparts of these curved cavities.

### 3.1 Cylindrical cavity

Figure 2 (a) shows the 3D drawing of the cylindrical cavity. The cavity is modeled with a ROB of 1 and ROC of 1, whilst the azimuthal angle $\phi$ is defined as $\phi = 0$ pointing along the positive direction of the x-axis. The calculated PSOS in Fig. 2 (b) indicates a completely symmetric WGM cavity, with all sinχ vs. $\phi$ being straight lines. This is not surprising when realizing that the cavity can be unfolded to a flat one, and indeed, the PSOS shares the same features as its unfolded flat counterpart (not shown here). Nevertheless, the curved cavity is obviously not symmetric in the DTG, as shown in Fig. 2 (c), exhibiting a sinusoidal variation of both the curvature and torsion of the light trajectory with $\phi$. It should be noted that in the DTG, light trajectories with various mode numbers $n$, i.e., light reflecting $n$ times at the cavity boundary within one round trip, are already included. On every point of a given segment of geodesic line between two adjacent reflecting points (regardless of $n$), the curvature and torsion are constants due to the nature of the helix. The value of $\phi$ for each segment is defined at its middle point, based on which the calculation shows identical distribution of curvature and torsion for all values of $n$, as shown in Fig. 2 (c).

The fact that the curvature and torsion of the light trajectory depend only on $\phi$ for all mode numbers induces anisotropic extra bending loss of all optical modes. Especially, at $\phi = \pm\frac{\pi}{2}$ both the curvature and torsion are zero, meaning a trajectory of a straight line, whilst at some other $\phi$ the curvature-induced extra bending loss can be much larger. This might yield a unique non-uniform loss mechanism depending on the integral effect of the overlap between light field and structural curvature.

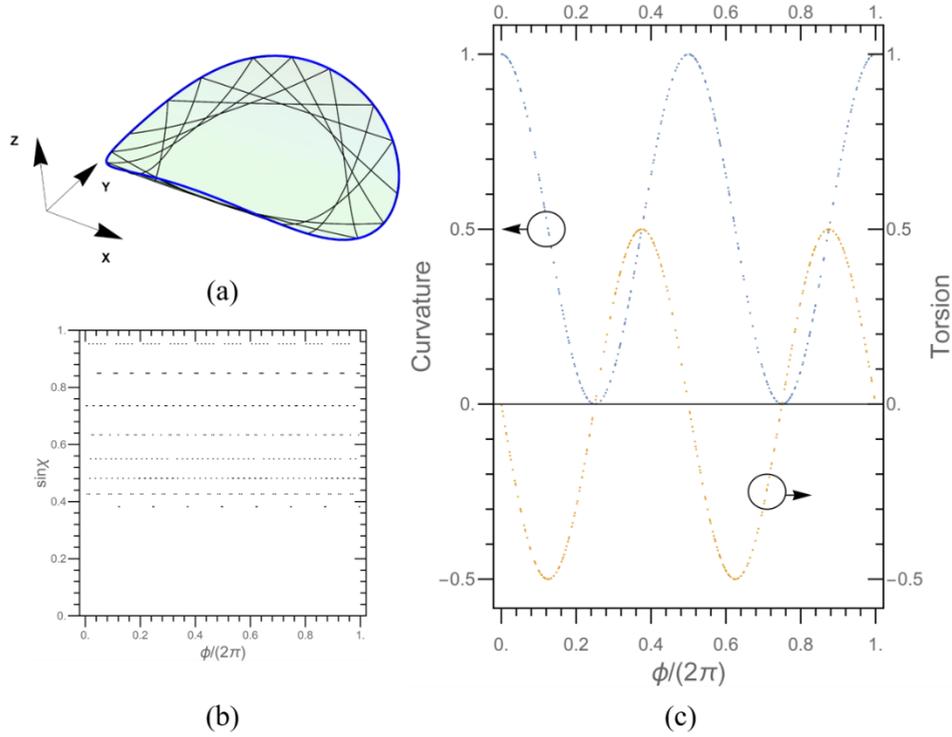

Fig. 2. Ray dynamics in a symmetric cylindrical cavity. (a) The 3D boundary of the cylindrical cavity, exhibiting a ROB of 1, and a ROC of 1. Black lines illustrate the calculated light trajectories. (b) The PSOS of the cavity, in which the sine of the reflection angle $\sin\chi$ is plotted as a function of the azimuthal angle $\varphi$ at the cavity boundary. (c) The DTG of the cavity, where the blue and yellow dots constitutes the curvature and torsion of the light's trajectory at each azimuthal angle $\phi$. In all graphs $\phi = 0$ is defined by the positive direction of the x-axis, and all lengths are in arbitrary units. The same for the rest of the figures in this article.

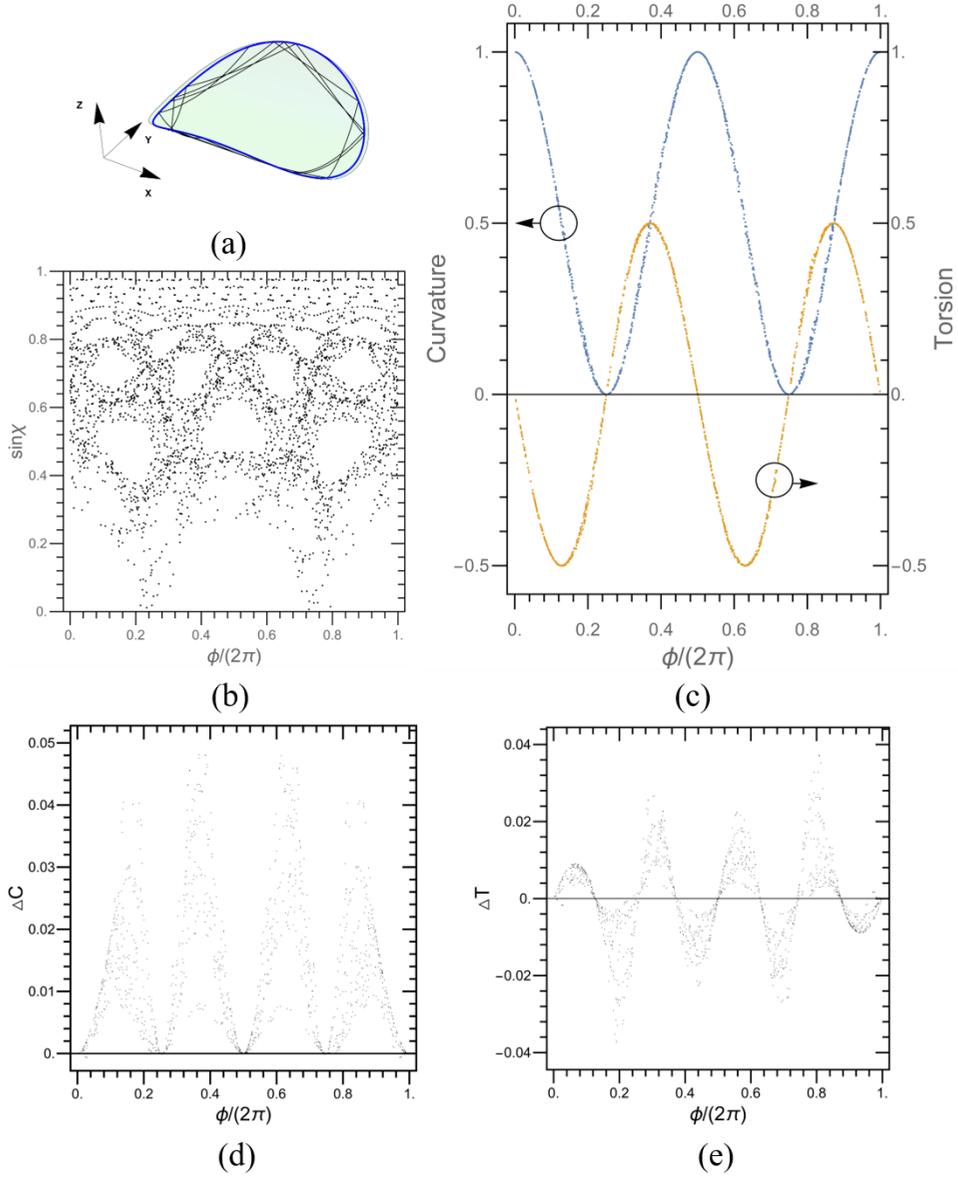

Fig. 3. Ray dynamics in a cylindrical Face cavity. (a) The 3D boundary of the cylindrical cavity, exhibiting a ROC of 1, and parameters of $a_2 = -0.1329, a_3 = 0.0948, b_2 = -0.0642, b_3 = -0.0224$ Black lines illustrate the calculated light trajectories. (b) The PSOS of the cavity. (c) The DTG of the cavity, where the blue and yellow dots constitutes the curvature and torsion. (d) and (e) The difference in curvature (d) and torsion (e) between the Face cavity and the symmetric cavity in Fig.2.

We further show that our algorithm can be used to calculate ray dynamics of curved cavities with deformed boundaries. Herein we set the well-know "Face cavity" as an example [16], but our model can be adapted to any arbitrary boundary as long as it can be expressed by an equation. The Face cavity on a cylindrical surface is shown in Fig. 3 (a), with its ROB defined by the following equation:

$$R = \begin{cases} R_0 \cdot (1 + a_2 \cdot \cos^2 \phi + a_3 \cdot \cos^3 \phi) & \phi \in \left(\dfrac{-\pi}{2}, \dfrac{\pi}{2}\right) \\ R_0 \cdot (1 + b_2 \cdot \cos^2 \phi + b_3 \cdot \cos^3 \phi) & \phi \in \left(\dfrac{\pi}{2}, \dfrac{3\pi}{2}\right) \end{cases} \quad (1)$$

In which $R$ is the ROB meaning the distance between the center and the boundary of the cavity along the geodesic lines of the cylindrical surface, with the parameters chosen to be $R_0 = 1, a_2 = -0.1329, a_3 = 0.0948, b_2 = -0.0642, b_3 = -0.0224$. The calculated PSOS and DTG are shown in Fig. 3 (b) and (c). Again, the result of PSOS just exhibits the same features as the one of a flat Face cavity (not shown here) as the cylindrical surface can be undoubtedly unfolded. On the other hand, the curvature and torsion of the light rays, which is a property purely from the curvature of the surface, shows similar sinusoidal features but with some variation from the cylindrical symmetric cavity of the same ROC, as presented in Fig. 3 (d) and (e). The differences, though being blow 5%, is anisotropic showing periodic fluctuations with $\phi$. Therefore, it is clear that the boundary geometry can have some tiny but traceable effects on the curvature and torsion of the light rays.

## 3.2 Spherical cavity

The shape and light trajectories of a spherical cavity with ROC=1 and ROB= π/3 is illustrated in Fig. 4 (a). Unlike the cylindrical cavity, the spherical surface is uniform and isotropic, leading to completely identical geodesic lines, i.e., the great circles of the sphere, exhibiting a constant curvature and zero torsion at every point. Nevertheless, the PSOS of the spherical cavity shows dramatic differences from its flat counterpart. As shown in Fig. 4 (b) whereas the shape of ROC=1 and ROB=π/6 is applied, the WGMs generally all shift upwards toward larger reflection angle sinχ, with the amount of shift decreasing with the mode number $n$. This is determined by the nature of the spherical surface on which the internal angle of a regular polygon is larger than that of its flat counterpart. It should be noted that $n$ can be non-integers in ray dynamics as light is considered as classical particles, whereas $n$ is defined as 2π divided by the central angle difference between two adjacent reflecting points. The upshifting trend is more clearly identified in Fig. 4 (c), where the sinχ is plotted as a function of ROB/ROC for a series of $n$. As the ray dynamics of the PSOS (but not the DTG) is only related to the shape but not the absolute size of the cavity, it is the value of ROB/ROC, i.e., the portion of the entire surface of a sphere that the cavity covers, instead of the absolute values of curvatures that matters. It is clearly seen from Fig. 4 (c) that the differences of sinχ among the WGMs of various $n$ decreases with increasing ROB/ROC. Particularly, at ROB/ROC=0 (ROC is infinite) the cavity reduces to its flat counterparts, and at the extreme case of ROB/ROC=π/2 where the cavity becomes a surface of hemisphere with its boundary itself being a great circle, the WGMs reaches an interesting regime where the trajectories of $n=2$ WGMs can exhibit arbitrary reflection angles and the trajectories of all the $n>2$ WGMs must be the cavity boundary itself, indicating a definite condition of sinχ = 1. Indeed, those $n>2$ modes seem to reduce to the WGMs in a sphere-shaped cavity in flat 3D space from an experimental point of view, which can be analyzed using straight-light ray dynamics. However, they are very different in terms of the mathematical model as the light ray has to be curved herein, which can be approximated by infinitely large mode number (or reflection angles of definite π/2) of the 3D sphere-shaped cavity where the straight-line trajectories can be considered to be infinitely close to the surface of the cavity. As the infinite case does not exist in reality, i.e., there is no way to couple light of π/2 refraction angle into the cavity, the model of spherically curved cavity should be restricted to the regime of ROB/ROC<π/2, indicating that the cavity should not contain a full great circle.

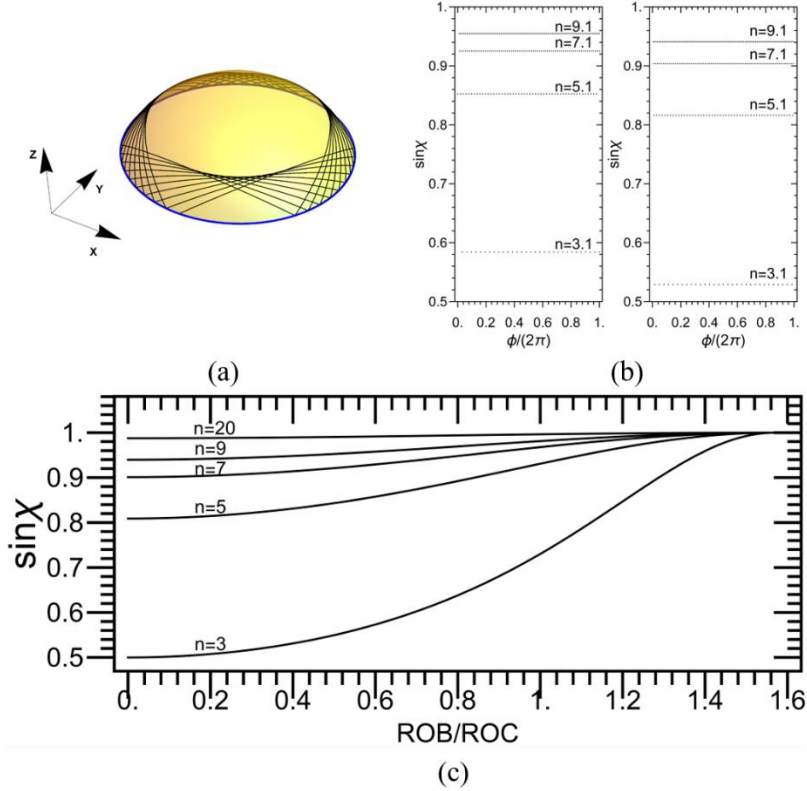

Fig. 4. Ray dynamics in a symmetrical spherical cavity. (a) The 3D boundary of the spherical cavity, exhibiting a ROC of 1, and a ROB of π/3. Black lines illustrate the calculated light trajectories. (b) The PSOS of a symmetrical spherical cavity of ROB=π/6, ROC=1 (left panel) and its flat counterpart (right panel) for a series of mode numbers $n$. (c) The light reflection angle as a function of $n$ for different values of ROB/ROC.

As what has been done for the cylindrical cavity, we perform ray dynamics analysis of a "Face cavity" defined on the spherical surface. The equation of the boundary is of the same expression as in Eq. (1), as presented in Fig. 5 (a). Herein the chaotic trajectories do not repeat that of the flat counterpart, as in differential geometry a continuous shape variation from a flat to curved surface generally does not keep the mapping of all straight lines in flat space to geodesic lines in curved space. Nevertheless, it is interesting that all the periodic features of the WGMs and islands remain (though perhaps with small shape differences), but shifted towards larger sinχ compared to the flat counterparts, as shown in Fig. 5 (b) and (c), demonstrating again an important feature from the nature of the positively curved surface.

Despite the demonstration of significant discrepancies between WGMs in flat and curved surfaces, more interesting applications will probably require a microcavity with non-uniformity of both ROB and ROC. One example of such applications would be the design of an out-of-plane unidirectional WGM laser: the deformation of boundaries leads to coexisting WGMs and chaotic modes that are linked by dynamical tunneling [16], while designed non-uniform space curvature yields local maxima of light-ray curvature which is on the trajectories of the chaotic modes but not of the WGMs. In such configurations the lasing occurs in the low-loss WGMs and dynamically tunnels to the chaotic modes of very large local curvature which renders locally high bending loss, resulting in a unidirectional emission from the cavity area of any designed orientation in 3D space, instead of from the cavity boundary as conventional WGM cavities. To realize such applications one will need to calculate geodesic lines of non-uniformly curved space, which would require sophisticated skills in differential geometry if analytical solutions exist for the designed shape, or numerical solutions to geodesic equations if not analytically solvable. It has been recently shown that conformal transformation optics can simplify the study of rotationally-symmetric 3D curved space by transforming it to a 2D gradient refractive index profile [35,36]. Nevertheless, for an arbitrarily curved space, one would need to further develop the method to work beyond the conformal regime.

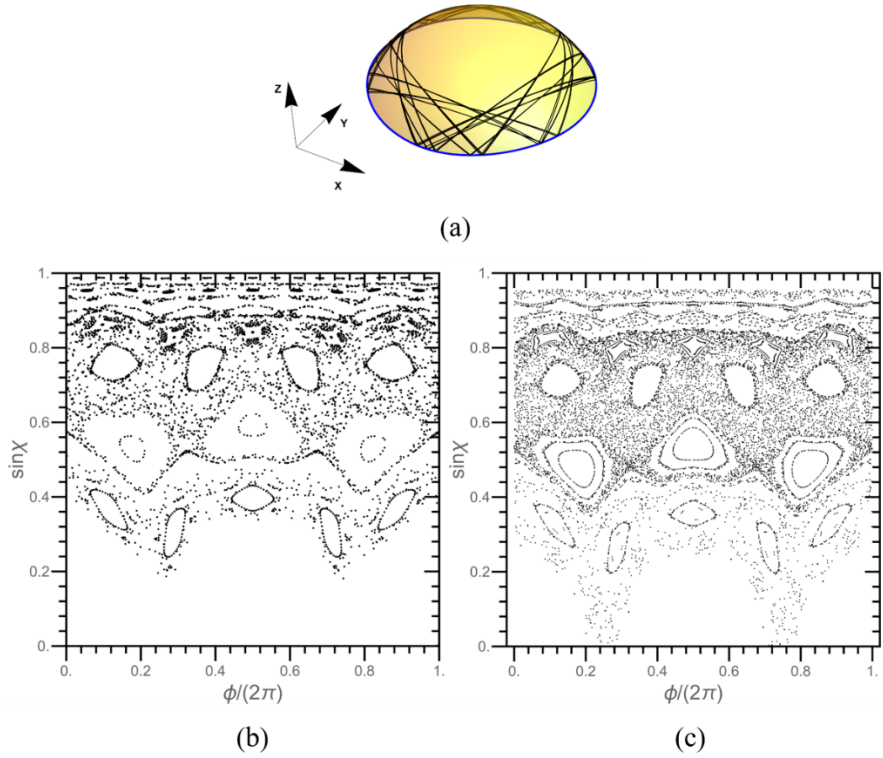

Fig. 5. Ray dynamics in a spherical Face cavity. (a) The 3D boundary of the spherical cavity, exhibiting a ROC of 1, and parameters of $R_0 = \pi/3, a_2 = -0.1329, a_3 = 0.0948, b_2 = -0.0642, b_3 = -0.0224$. Black lines illustrate the calculated light trajectories. (b) The PSOS of the spherical Face cavity with $R_0=\pi/6$, all other parameters same as in (a). (c) PSOS of a flat Face cavity with the same parameters as in (b).

## 4. Conclusion and Perspectives

In conclusion, we proposed a novel method of ray dynamics for microcavities defined on curved space, and demonstrated dramatic differences of the WGMs between cavities on curved surface and their flat counterparts, reflecting the nature of spatial curvature. The model should apply in a situation that both the ROC and ROB of the 2D surface are much larger than the optical wavelength. Although in this Article we chose regularly-curved surface whose geodesic equations are known for simplicity, the proposed geodesic ray-tracing method is, in principle, applicable for more complicated curved spaces provided suitable mathematical tools can be further developed. Experimental realization of curved cavities could be achieved via material strain, sophisticated lithography, laser milling and high resolution 3D printing, etc., though reaching a precise control of curvature on microscale remains challenging. With the waveguide properties of the actual curved surface taken into consideration, curved space-time can be achieved by varying local thicknesses of the membrane, which leads to variation of the effective refractive index and thereby the effective metric of time [37]. The cavity defined on curved and thickness-varying waveguides would eventually provide a promising platform for optical simulators of quantum effects in gravity [38,39].

## 5. Acknowledgement

This work is supported by National Key R&D Program of China (2017YFA0303703), and National Natural Science Foundation of China (12074303, 11804267 and 11904279).

**Disclosures**
The authors declare no conflicts of interest.